\documentclass[onecolumn]{openjournal}

\usepackage{lineno}

\usepackage{amssymb,amsmath,bm}
\usepackage{mathtools}
\usepackage{xcolor}
\usepackage{etoolbox}
\usepackage[colorlinks=true,linkcolor=blue,citecolor=blue,urlcolor=blue]{hyperref}

\makeatletter

\def\frontmatter@title@below{%
  \vspace*{-2.63\baselineskip}%
  \vspace*{0.25in}%
}

\pretocmd{\@sect}{\def\@currentcounter{#1}}{}{%
  \PackageWarning{ojoa-hyperref}{Could not prepend current counter to \string\@sect}%
}

\makeatother

\newcommand{\nv}{\hat{\bm n}}
\newcommand{\deq}{\coloneqq}
\newcommand{\cN}{\mathcal{N}}
\newcommand{\cS}{C}

\newcommand{\de}{\mathrm{d}}
\newcommand{\fsky}{f_\mathrm{sky}}

\def\dd{ \text{d} }
\def\la{\langle}
\def\ra{\rangle}

\begin{document}

\title{The impact of the formation channel on gravitational-wave--galaxy cross-correlations}
\shorttitle{GW--galaxy cross-correlations}

\author{Kabir Chakravarti$^{1,2,\star}$}
\author{Federico R.~Urban$^2$}

\affiliation{$^1$Chennai Mathematical Institute, Plot H1 SIPCOT IT Park, Siruseri 603103, Tamil Nadu, India}
\affiliation{$^2$CEICO -- FZU, Institute of Physics of the Czech Academy of Sciences, Na Slovance 1999/2, 182 00 Prague, Czech Republic}

\thanks{$^\star$E-mail: \href{mailto:chakravarti@fzu.cz}{chakravarti@fzu.cz}}
\shortauthors{Chakravarti and Urban}

\begin{abstract}
The angular, harmonic cross-correlation between gravitational wave (GW) events and galaxy catalogues contains rich information on the large-scale structure and the origin of compact binary mergers. In this work, we study how uncertainties in the binary formation channel affect the predicted cross-correlation signal for both current-generation and next-generation networks of detectors. We generate five mock GW catalogues for which we vary the progenitor-to-remnant mass-transfer function and the time-delay probability distribution between progenitor and remnant. We then cross-correlate these catalogues with galaxy samples modelled on the 2MASS Photometric Redshift catalogue (2MPZ) and the Gaia-unWISE quasar catalogue (Quaia). We find that the mass-transfer function has negligible effect on the cross-correlation signal, with differences remaining within redshift uncertainties. In contrast, the time-delay distribution dramatically affects the redshift distribution of the GW events and, with it, the cross-correlation signal, particularly for shallow galaxy catalogues. In particular, current-generation facilities can achieve significant detections only for the longest time delays when cross-correlated with 2MPZ, whilst all cross-correlations with the deeper Quaia catalogue are marginally detectable or consistent with zero. Our exploratory results thus demonstrate that forecasts on cosmological or astrophysical parameters derived from GW-galaxy cross-correlations are, as expected, strongly sensitive to the assumed binary formation history.
\end{abstract}

\maketitle

\section{Introduction}
\label{sec:intro}

The direct observation of dozens of gravitational wave (GW) events has opened a new window on the study of the Universe. The waveforms of GW events from the merger of two black holes (BHBH), two neutron stars (NSNS) or a black hole and a neutron star (BHNS) contain rich information about the physical properties of the merging objects, including their masses, spins, and orbital characteristics (see, e.g.~\citet{LIGOScientific:2016aoc, LIGOScientific:2017vwq, Mandel:2021ewy}). Collectively, catalogues of GW events~\citep{KAGRA:2021vkt} inform us about the properties of both the compact objects' progenitors and the environments in which the mergers take place, which help us understand the history of the early Universe. For instance, it is expected that the spatial distribution of mergers follows that of the large-scale structure (LSS), and that the precise way this correlation is realised can tell us, for example, whether merging BHs have a primordial or astrophysical origin~\citep{Raccanelli:2016cud}. Indeed, in the first case GWs from BHBH mergers are seeded by black holes born before the formation of galaxies and therefore track closely the distribution of cosmological dark matter -- and would constitute a fraction of it; in the second case the GW events follow more closely the distribution of visible matter, which is a biassed tracer of the LSS.

A powerful tool to address this and other cosmological questions is the harmonic, angular cross-correlation (XC) between GW and galaxy catalogues~\citep{Oguri:2016dgk, Scelfo:2018sny, Calore:2020bpd, Libanore:2020fim, Libanore:2023ovr, Ferri:2024amc, Pedrotti:2025tfg, SantiagodeMatos:2025iyj}. At present, because the number of observed GW events is relatively small, have a very limited redshift reach and poor angular sky localisation, such analyses have a minimal statistical power. However, in the future, with the advent of the next generation of both ground-based and satellite GW detectors such as the Einstein Telescope (ET), Cosmic Explorer (CE) and LISA, we expect to detect several orders of magnitude more GW events. These events will reach much farther in distance and with much better sky localisation, especially when a network of detectors is considered, thereby significantly improving the statistical power of this messenger.

A key ingredient in the forecast of the ability of future GW catalogues to accurately determine cosmological or astrophysical parameters is a realistic estimate of the redshift and signal-to-noise-ratio (SNR) distributions for a population of GW events. These distributions directly influence the observed GW merger rate that a given detector, or a combination of such detectors, will observe within a given data campaign, and also determine how well these events can be localised in the sky. These two distributions depend on the cosmological history of the population of merging binaries, namely how many binaries with given masses (and spins) merge and at which redshift. As an example,~\citet{Calore:2020bpd, Libanore:2023ovr} forecast the ability to distinguish between different merger origins (primordial or astrophysical) based on the determination of the GW bias \(b\) with respect to the LSS, which quantifies the degree to which the GW source distribution follows the distribution of matter in the Universe. If the details of the merger history change, the expected XC power spectra change, and so will the estimates on the determination of the bias.

In this work, we take a closer look at some of the parameters that go into the determination of the GW merger rate, in order to assess their impact on the XC. In particular, we look at how the redshift distribution of the GW events (which we obtain by simulating GW mergers starting from a given star formation rate) is modified if we change: (a) the remnant-mass transfer function, which describes the mass of the remnant at the endpoint of stellar evolution of its progenitor star, and (b) the probability distribution function for the time delay between progenitor and remnant. In order to do so we will follow closely the formalism of~\citet{Calore:2020bpd} and assume a background cosmological model which will enable us to convert luminosity distance measurements for GW events into redshift estimates -- in doing so we will ignore subdominant lensing and relativistic effects which could be relevant at much smaller angular scales than those we survey here.

This paper is organised as follows. In the next \autoref{Sec:form} we will introduce our formalism for the XC. We describe the GW and galaxy catalogues in \autoref{Sec:cats}. We then present our results in \autoref{Sec:res} and conclude in \autoref{Sec:out}.

\section{The formalism}\label{Sec:form}

Any function \(\Delta_a(\nv)\) of direction \(\nv\) on the celestial sphere can be decomposed into harmonic coefficients as
\begin{align}\label{eq:sph_harm}
	\Delta_{\ell m}^a &\deq \int\dd\nv\,Y^*_{\ell m}(\nv)\,\Delta_a(\nv) \;,
\end{align}
where \(Y_{\ell m}\) are Laplace's spherical harmonics. In this work the functions being correlated are the anisotropy fields of galaxies and GWs, for which \(a\in\left\{\mathrm{g},\mathrm{GW}\right\}\), respectively. For each catalogue we write the anisotropic field on the sphere as
\begin{align}\label{eq:ani}
    \Delta_a(\nv) &\deq \int \de\chi\;\phi_\mathrm{a}(\chi)\,\delta_a(z,\chi\,\nv) \;,
\end{align}
where \(\delta_a(z,\chi\,\nv)\) is the three-dimensional overdensity of GWs or galaxies. The radial kernel \(\phi_a(\chi)\) (normalised to 1 along the line of sight) represents the distribution of objects (GW events or galaxies) as a function of radial comoving distance, \(\chi\), which is related to the cosmological redshift \(z\) through \(\de\chi/\de z=1/H(z)\), with \(H\) the Hubble parameter. We model both overdensities as linearly biassed tracers of the LSS, namely \(\delta_a \deq b_a \delta_\mathrm{LSS}\). Since we are not looking to determine the galaxy bias we consider it to be a fixed quantity which we model as inversely proportional to the growth factor $D(z)$ for a given cosmological model. As regards the GW bias \(b_\mathrm{GW} \deq b\), in principle this is a \(z\)-dependent and scale-dependent quantity. However, our goal is not to determine the bias but rather to study how the XC changes given different binary formation histories; therefore, even keeping in mind that a strong redshift dependence could impact our reconstructed GW radial kernels, for simplicity in this work we choose to consider this parameter as a constant overall multiplicative factor for the XC, defined as \(\delta_\mathrm{GW}\deq b\delta_\mathrm{LSS}\).

Our main observable is the harmonic, angular power spectrum \(\cS_\ell^{ab}\) between two projected quantities \(\Delta_a\) and \(\Delta_b\), which, for broad kernels such as those we work with here, is defined as the two-point function of the harmonic expansion coefficients of the anisotropy fields i.e.\
\begin{align}\label{eq:aps}
	\langle \Delta^a_{\ell m}\,\Delta^{b\ast}_{\ell'm'}\rangle &\deq \delta_{\ell\ell'}\,\delta_{mm'}\,\cS^{ab}_\ell \;,
\end{align}
where the angle brackets stand for the ensemble average. When the two fields differ \(a \neq b\) we speak of the angular cross-correlation power spectrum, whereas the auto-correlation spectrum is for \(a=b\).\footnote{Notice that whenever we speak simply of ``spectrum'' we always mean the angular power spectrum, not the matter power spectrum of Eq.~\eqref{eq:cl_limber}.} The spectrum \(\cS_\ell^{ab}\) is most conveniently expressed in terms of the three-dimensional matter power spectrum \(P(z,k)\) as
\begin{equation}\label{eq:cl_limber}
	\cS^{ab}_\ell= b b_\mathrm{g}\int \frac{\de\chi}{\chi^2}\;\phi_a(\chi)\,\phi_b(\chi)\,P\left[z(\chi),k=\frac{\ell+1/2}{\chi}\right] \;.
\end{equation}
We obtain the matter power spectrum \(P(z,k)\) with the python implementation of the Core Cosmology Library \texttt{pyccl}.\footnote{\url{https://github.com/LSSTDESC/CCL}} We assume standard $\Lambda$CDM cosmology with parameters $\Omega_c=0.25$, $\Omega_b = 0.05$, $\Omega_k = 0$, $\sigma_8 = 0.81$, $n_s = 0.96$, $h = 0.67$ and no massive neutrinos.

The error on the XC signal for a simple noise-dominated process can be estimated analytically in terms of the number of objects, and the field of view and angular resolution in each catalogue:
\begin{equation}
\label{eq:cross}
\left(\frac{\delta C_\ell^{ab}}{\cS_\ell^{ab}}\right)^2=\frac{1}{(2\ell+1)\Delta \ell \fsky}
\left[ 1+\frac{\cS_\ell^{aa}\cS_\ell^{bb}}{(\cS_\ell^{ab})^2}\left(1+\frac{\cN_\ell^{aa}}{{\cal B}_{\ell,a}^2\,\cS_\ell^{aa}}\right)\left(1+\frac{\cN_\ell^{bb}}{{\cal B}_{\ell,b}^2\,\cS_\ell^{bb}}\right) \right]\, ,
\end{equation}
where $\Delta\ell$ is the bandpower width and \(\fsky\) is the fraction of the sky covered by the catalogue with the smallest coverage of the two. Since the angular positions of the GWs and the galaxies are discrete point processes, we can write the auto-correlation noise terms as
\begin{equation}\label{eq:noise}
	\cN^{aa}_\ell=\frac{4\pi \fsky}{\bar{N}_{\Omega,a}} \;,
\end{equation}
where \(\bar{N}_{\Omega,a}\) is the angular number density of points in sample \(a\).

The last ingredient in Eq.~\eqref{eq:cross} is the angular resolution of each event, namely the point-spread function or beam, which is accounted for by a circular angular window function ${\cal B}_{\ell,i}$ with a Gaussian profile with width \(\sigma\). Because galaxy catalogues have a much better resolution than the sky localisation of GW events, in what follows we ignore the beam for the galaxy catalogues. The procedure for the calculation of the beam for our GW maps is described in detail in \autoref{Sec:beam}.

\section{The catalogues}\label{Sec:cats}

\subsection{Gravitational Waves}\label{Sec:MK}

In order to understand the details of the cosmological merger history, namely how isolated progenitor stars become trapped in binaries, evolve and eventually merge, we perform a population synthesis by running simulations of a set of progenitors to compact binaries, assuming a specific {\it formation channel}. The details of the GW catalogue that arises from the population synthesis strongly depend upon the history of the generated binaries. Changing the astrophysical parameters that determine the formation channel introduces a complex chain of changes which affect the GW event detection rate (see, e.g.~\citet{Santoliquido:2020bry, Santoliquido:2020axb}), and with it the XC.

In all our simulations the formation channel follows the prescriptions of~\citet{Belczynski:2016jno,Belczynski:2016obo}. Specifically, we assume that compact binaries primarily form through the evolution of isolated binary star systems, where two main-sequence stars remain gravitationally bound as they evolve and eventually become compact object remnants, provided the binary system survives the various evolutionary stages. This assumption disregards the possibility of binary formation through dynamical interactions, including the capture of isolated remnants, which however can be significant in dense environments such as globular clusters~\citep{Hong:2018bqs}, young star clusters~\citep{Torniamenti:2022txt} and active galactic nuclei~\citep{Grobner:2020drr}, see also~\citet{Stevenson:2022djs}. As is standard in the literature, we also assume that the main sequence stars are born obeying the Madau-Dickinson star formation rate (SFR)~\citep{Madau:2014bja} and an initial mass function (IMF) given by~\citet{Kroupa:2000iv}. Additionally, we assume that only a fraction of stars at any redshift end their lives as merging compact binaries, with the trapping factor \(f_b\) representing the fraction of binaries that successfully form and merge despite supernova natal kicks imparted to the compact objects during their formation~\citep{deFreitasPacheco:2005ub}. We adopt the minimum delay times between the formation of progenitor stars and the merger of their compact remnants as proposed in~\citet{Calore:2020bpd}: we use 50~Myr for binary black hole (BHBH) systems, 20~Myr for binary neutron star (NSNS) systems, and 30~Myr for black hole-neutron star (BHNS) systems. Lastly, in generating the binaries, for BHs we randomly assign spins with a uniform distribution between~0 and~1 that are aligned or anti-aligned with respect to the orbital angular momentum; the spins of the NSs are all set to zero.

Once the GWs are produced at the source, whether these will be detected depends on the response of the (network of) active detectors, which is determined by its sensitivity and antenna pattern. For our purpose we use a current-generation network consisting of the LIGO-H, LIGO-L and VIRGO detectors (HLV) as well as a next-generation network consisting of the two Cosmic Explorers at the Hanford and Livingston locations and the Einstein Telescope at the VIRGO location (ET2CE). The directions of the simulated GW events have been chosen randomly from a uniform distribution over the sphere. The chances of detection and localisation of each individual event depends on its actual position, owing to the explicit angular dependence of the antenna pattern function of the detector network. However, for sufficiently large gravitational wave catalogues and multiple simulations of a given population, the specific sky directions of individual events become statistically insignificant. This is because the distribution of SNRs for the events becomes well-sampled -- in particular, for our expected current-generation detection rates of \(600/\mathrm{yr}\) (see below), we expect the shot noise related to the sky distribution to contribute only a few percent to the overall XC. Lastly, notice that, while clearly beneficial for this type of study, we conservatively do not make use of the fact that some GW events may have a detected electromagnetic counterpart, because significant conclusions about the formation channel can be obtained without them by relying solely on the direct GW distance measurement.

We are now interested in quantifying how the GW merger rate changes when we look more closely at some of the astrophysical parameters that determine the dynamics of the formation channel. In particular, we study the impact of (a) the progenitor-to-remnant mass-transfer function and (b) the probability distribution function for the time delay \(\tau\) between progenitor and remnant. In addition to that, we study the effect of changes in the IMF and the binary trapping factor \(f_b\). We describe these in more detail below.

First, concerning the progenitor-to-remnant mass-transfer function, we generated different GW catalogues corresponding to the ``rapid'' and ``delayed'' nova engines as described in~\citet{Fryer_2012} with stellar metallicity fixed to the Sun's metallicity. Our analysis indicates that incorporating redshift-dependent stellar metallicity evolution has a minimal direct impact on the remnant mass function. However, the metallicity and its cosmological evolution affects other aspects of stellar and binary evolution~\citep{Boco:2020pgp, Scelfo:2020jyw, Pellouin:2024qao}; in order to keep the problem tractable, in this first work we do not pursue this further and, despite the fact that the metallicity can significantly change the evolution history of the mergers, we factor these uncertainties into the description of the IMF, see below.

The second parameter that we explore is the delay function, namely the probability function assumed for estimating the coalescence time, \(P(\tau)\). This is typically obtained from $n$-body simulations such as those performed in~\citet{deFreitasPacheco:2005ub} and it depends on the presence of specific dynamic processes like the Roche lobe overflow and a common envelope phase, as well as the presence and dynamics of the stars' natal kicks. Performing such $n$-body simulations is beyond the scope of this first work; therefore, in order to study the importance of the delay function on the formation channel, we choose to treat it as a free function. We parametrise the delay function as a power law, $P(\tau) \deq 1/{\tau^\alpha}$ and set the power $\alpha = 1$ to be our baseline case. We will then compare it to three other representative values of $\alpha$, namely $\alpha = 0.95$, $\alpha = 0.75$ and $\alpha = 0.50$ in order to bracket different realistic scenarios~\citep{Mukherjee:2021qam, Pellouin:2024qao}; furthermore, it has recently also been suggested that the time-delay distribution for observed binary black holes is not universal, see~\citet{Afroz:2025typ} (a possibility which we leave for future work). Notice that what we refer to as formation channel here is a given choice of model and parameters that describe the evolution history (formation and merger) of the binary, starting from the IMF and SFR, including the trapping factor, the progenitor-to-remnant mass-transfer function and the probability distribution function for the time delay between progenitor and remnant. The dynamics of a common envelope evolution and dynamical capture are, in our case, described by the delay function, which we treat as a free parameter -- in other words, if there are deviations from the standard common envelope evolution, the slope of the delay function will reflect this.

A third astrophysical parameter that describes the compact binary formation channel is the trapping factor. The trapping factor appears as an overall multiplicative factor that represents the efficiency with which two compact objects become gravitationally bound, and is also connected to the fraction of compact binaries that survive a natal kick to the remnant~\citep{Santoliquido:2020bry, Santoliquido:2020axb, deFreitasPacheco:2005ub}. Because of the complex dynamical processes involved in the formation of the compact binary, the trapping factor \(f_b\) is poorly understood and not very well measured. For instance,~\citet{Belczynski:2016jno, Belczynski:2016obo} argue in favour of a complete absence of novae anywhere in the formation channel because novae tend to disrupt binaries and reduce detection rates below those seen by GW detectors.

A fourth ingredient in the determination of GW merger rates is the SFR. The uncertainties in the SFR are of order \(10\%\) for low-redshift and can increase up to order~\(1\) for redshifts beyond cosmic noon~\citep{Madau:2014bja}. These uncertainties mostly affect the number and mass of high-redshift progenitors which, depending on the progenitor-remnant time delays, can cascade onto the observed rates. In our specific case we do not expect this to have a significant impact on cross-correlations with shallow galaxy catalogues (see \autoref{Sec:gal}), but it can change the cross-correlations with deeper galaxy catalogues. However, for simplicity here we choose to keep the SFR fixed to the best fit of~\citet{Madau:2014bja}.

Lastly, a further poorly understood component is the IMF. Currently our knowledge about the IMF is derived by observing the stellar distribution in the solar neighbourhood, the galactic disc and bulge and in different stellar clusters of our Galaxy~\citep{Kroupa:2000iv, Kroupa:2002ky}. Different models predict significantly different IMFs depending on the observation (see~\citet{Kroupa:2002ky, Kroupa:2024nax}). Moreover, there is evidence towards a non-negligible evolution of the IMF across redshift~\citep{Dave:2007km, vanDokkum:2007dy}. The effect of the different models for the IMF on the final merger rate is complex and, lacking an obviously preferred model for it, and in order to keep the number of parameters tractable, we choose to parametrise this uncertainty into an overall multiplicative factor in the final merger rate, which we fix in order to match the expected rates for a given set of detectors.

To summarise, we have four model parameters that have an effect on the merger detection rates: the mass-transfer function (``rapid'' or ``delayed''), the delay function $1/{\tau^\alpha}$ (with $\alpha = 1$, $\alpha = 0.95$, $\alpha = 0.75$ and $\alpha = 0.50$), the trapping factor and the IMF model. The trapping factor simply contributes an overall multiplicative factor in the merger rate, and, due to the complexity of the IMF modelling, we also choose to pack its uncertainty together with the trapping factor into a single constant $f_\mathrm{Res}$, which we fix by ensuring that the expected HLV detection rate is consistent with the expected value of $600/\mathrm{yr}$ for a fifth-observing-round (O5) sensitivity~\citep{Baibhav:2019gxm, Kiendrebeogo:2023hzf}. We summarise the properties of our synthetic GW catalogues alongside their event normalisation factor $f_\mathrm{Res}$ and the predicted next-generation merger rates in \autoref{tab:cats}. In order to compute the detection rates we have assumed the detection threshold SNR values of $\rho\geq8$ for HLV and $\rho\geq12$ for ET2CE; this is justified because events with very low SNR are very poorly localised and would not contribute to the XC, so we can discard them. Notice that, while the normalisation $f_\mathrm{Res}$ sets the current-generation detection rate for all our mock catalogues to be the same, because the mass-transfer function and the time delay change the redshift history of the mergers, $f_\mathrm{Res}$ does not force the next-generation detection rates to be the same. In other words, the ratio of next-generation to current-generation detections is unaffected by this normalisation.

\begin{table}
\centering
\begin{tabular}{|c|p{2.5cm}|c|c|c|} 
\hline
Catalogue ID & Mass Transfer  & $\alpha$ & $\langle n\rangle_{\mathrm{et2ce}}$ & $10^5 f_{\mathrm{Res}}$ \\ [0.3ex]
\hline
\phantom{0}6 & Rapid Nova   & 1.00 & 79972 & 14.634 \\ 
\hline
\phantom{0}7 & Delayed Nova & 1.00 & 81889 & 13.636 \\
\hline
\phantom{0}8 & Rapid Nova   & 0.95 & 45078 & \phantom{0}8.000 \\
\hline
\phantom{0}9 & Rapid Nova   & 0.75 & \phantom{0}7985 & \phantom{0}1.000 \\
\hline
11 & Rapid Nova & 0.50 & \phantom{0}3908 & \phantom{0}0.307 \\
\hline
\end{tabular}
\caption{GW catalogues along with their mass-transfer functions, delay function exponent $\alpha$, rescaling factor $f_\mathrm{Res}$ and corresponding population-averaged next-generation detection rate $\langle n\rangle_\mathrm{et2ce}$.}
\label{tab:cats}
\end{table}

\begin{figure}[htbp]
    \centering
    \includegraphics[width=\linewidth]{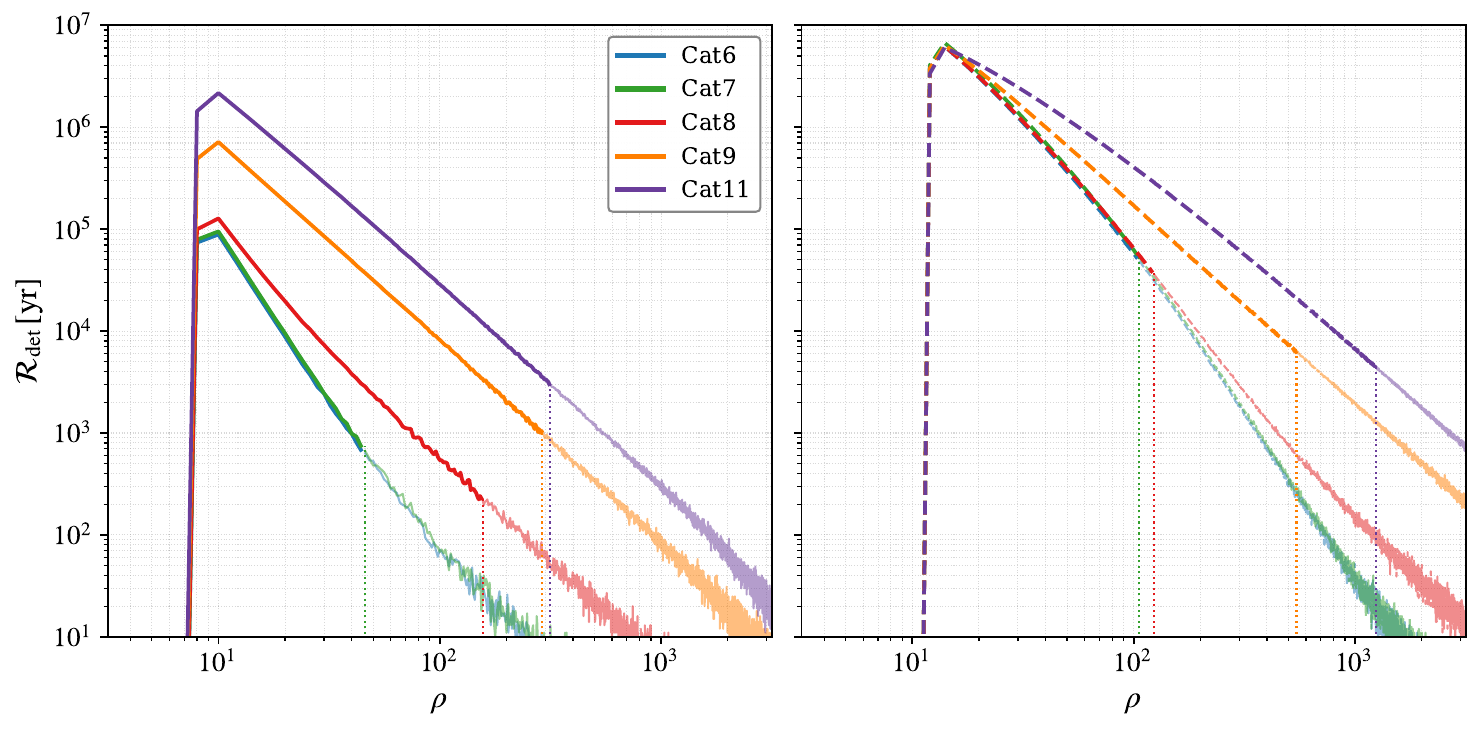}
    \caption{Distribution of SNRs $\rho$ with the current-generation ($\rho\geq8$, left, solid) and the next-generation ($\rho\geq12$, right, dashed) detectors for our catalogues: catalogue~6 is in blue, catalogue~7 in green, catalogue~8 in red, catalogue~9 in orange, and catalogue~11 in purple. The vertical dotted lines represent the $97\%$ containment values of the distributions. The catalogues as shown here are not rescaled by $f_\mathrm{Res}$.}
    \label{fig:rhodets}
\end{figure}

In \autoref{fig:rhodets} we show the raw (i.e.\ not rescaled by $f_\mathrm{Res}$) distribution of the SNRs $\rho$ for the current-generation ($\rho\geq8$, left, solid) and the next-generation ($\rho\geq12$, right, dashed) detectors for our five catalogues. The SNR histograms are monotonically decreasing as expected. Additionally, we also observe that the histograms lose compact support at the high $\rho$ end. These outlier events can introduce spurious numerical artifacts and bias the statistics of our sample populations; we therefore cut our catalogues and discard events outside of the $97\%$ boundary, which are shown by the vertical dashed lines in \autoref{fig:rhodets}. We have verified that, while varying the containment level from \(95\%\) to \(98\%\) (with \(97\%\) as our baseline) does have an appreciable effect on the the cross-correlations -- and their detectability -- these changes are less significant than those induced in going from \(\alpha=1.00\) to \(\alpha=0.95\) (corresponding to catalogues~6 and~8).

Before moving on to the study of the properties of the resulting five mock GW catalogues, we want to emphasise that, in constructing these catalogues, we first fix all the relevant parameters (IMF, SFR, time delay, etc) and generate one set of mergers which gives us a single realisation of the population of mergers. We then repeat this procedure $15$~times for an ensemble of $15$~realisations of the population. Averaging over the realisations ensures that we densely sample the redshift and SNR distributions that characterise the merger populations.\footnote{Note that the $97\%$ boundary in the SNR distribution is calculated from the total of $15$~realisations.} We then proceed to rescale the total number of mergers in order to meet the requirement of $600/\mathrm{yr}$ events for current-generation rates, which in turn sets the expected rate for next-generation detectors.\footnote{Notice that, from this procedure, the only quantity that we use in the analysis is the GW radial kernel which we sample with the 15~realisations of a population; we also need the overall number of events in order to estimate the error on the XC. Besides these, the remaining ingredients for the cross-correlation come from theory (i.e.\ from the matter power spectrum), so that the individual GW events are not needed. Lastly, notice that, once we rescale the raw number of GW events to $600/\mathrm{yr}$ for current-generation detectors, the same rescaling factor must be used for next-generation detectors, because otherwise different formation channels would be normalised in different ways.}

\subsubsection{GW radial kernels}\label{Sec:ker}

The normalised redshift distribution of the GW events is the radial kernel \(\phi_\mathrm{GW}(\chi)\). We computed the redshift of a given GW event directly from the luminosity distance as inferred from the waveform (recall that we assume no electromagnetic counterparts), and neglecting the small contributions from relativistic and lensing effects. In order to compute the errors $\delta z$ on the GW redshift distribution, we made a pilot GW catalogue of $10$ events drawn from the total simulated sample for catalogue~6, excluding the events in catalogue~6 themselves, and ran a Bayesian inference on those events. The inference runs were performed by making use of the $\texttt{dynesty}$ sampler to sample the GW log-likelihood functions with the $\texttt{bilby}$ library.\footnote{\url{https://github.com/joshspeagle/dynesty}, \url{https://github.com/bilby-dev/bilby}} Notice that as we ran these inferences we fixed the priors of selected extrinsic parameters, namely the sky position and angle of orientation of the binaries, as delta functions centred upon their injected values~\citep{Pankow:2016udj}. This reduces our computational cost per simulation and thereby allows us to perform the necessary runs to create the pilot catalogue.\footnote{For current-generation detectors with relatively poor sky localisation the sky position uncertainty and the luminosity distance uncertainty are correlated because the antenna pattern functions (and therefore the detector response) can vary significantly across the reconstructed GW sky patch. Because, as we shall see, the differences in radial kernel induced by even small changes in the progenitor-to-remnant time-delay is much larger than the redshift error this correlation could induce, we ignore the latter in what follows.}

\begin{figure}[htbp]
    \centering
    \includegraphics[width=\linewidth]{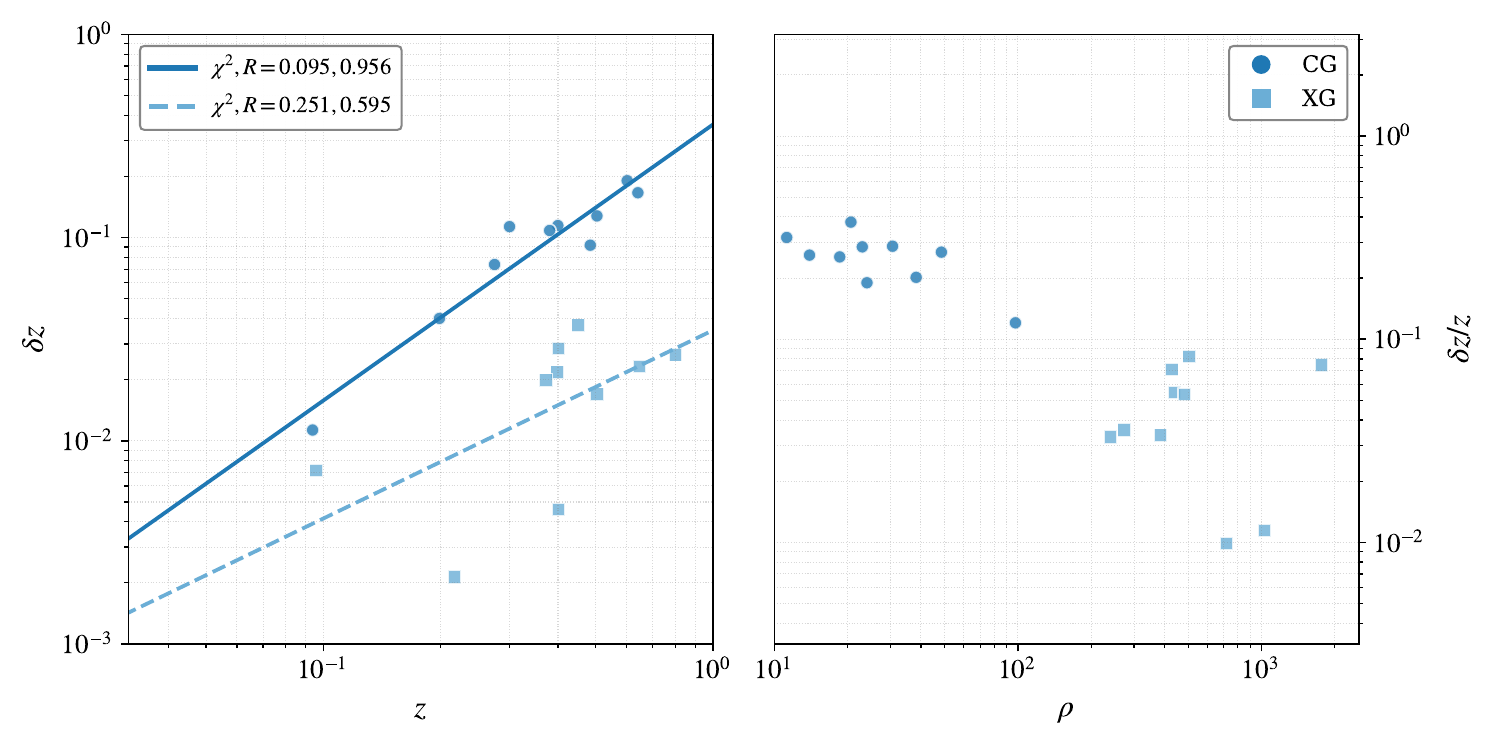}
    \caption{Scaling of the redshift errors $\delta z$ with redshift $z$ (left panel) and the SNR $\rho$ (right panel) for a pilot catalogue of $10$~events with current-generation (dark blue disks) and next-generation networks (light blue squares). The solid and dashed lines shows the best fits for these data points, respectively.}
    \label{fig:zerrs}
\end{figure}

\autoref{fig:zerrs}, left panel, shows the results of these inferences for current-generation (dark blue disks) and next-generation (light blue squares) detectors. In \autoref{fig:zerrs}, right panel, we show instead the relationship between the redshift errors and the SNR \(\rho\). In these figures we also show the best linear fits obtained for the scaling relations of $\log\delta z$ with $\log z$ as $\log\delta z = a \log z + b$, with $a = 1.36$, $b = -0.44$ and $a = 0.93$, $b = -1.46$ for current-generation and next-generation facilities, respectively. We use our $\delta z(z)$ relations to calculate the redshift errors associated with catalogue~6, in other words, for each event for which the redshift is estimated from the luminosity distance, we associate a redshift error according to our $\delta z(z)$ relation.\footnote{As can be seen from the \(\chi^2\) results, the fit is not very good for next-generation events; however, lacking any clear better expected fitting function, being limited in pilot catalogue size by computational costs, and given that our recovered redshift errors are in general agreement with the literature, we keep this as our working fit. We also further stress that we limit ourselves to only 10~events for both current-generation and next-generation catalogues because of the computational cost of running the inferences in which at least 6~intrinsic parameters need to be simultaneously varied for every event, in order to have an unbiassed estimation of \(\delta z\); finally, the error on \(z\) is the only quantity we obtain from this pilot catalogue.} The scaling with the SNR \(\rho\) verifies that our values of redshift errors and the corresponding fits are compatible to corresponding estimates obtained in~\citet{Calore:2020bpd}. In 

\begin{figure}[htbp]
    \centering
    \includegraphics[width=\linewidth]{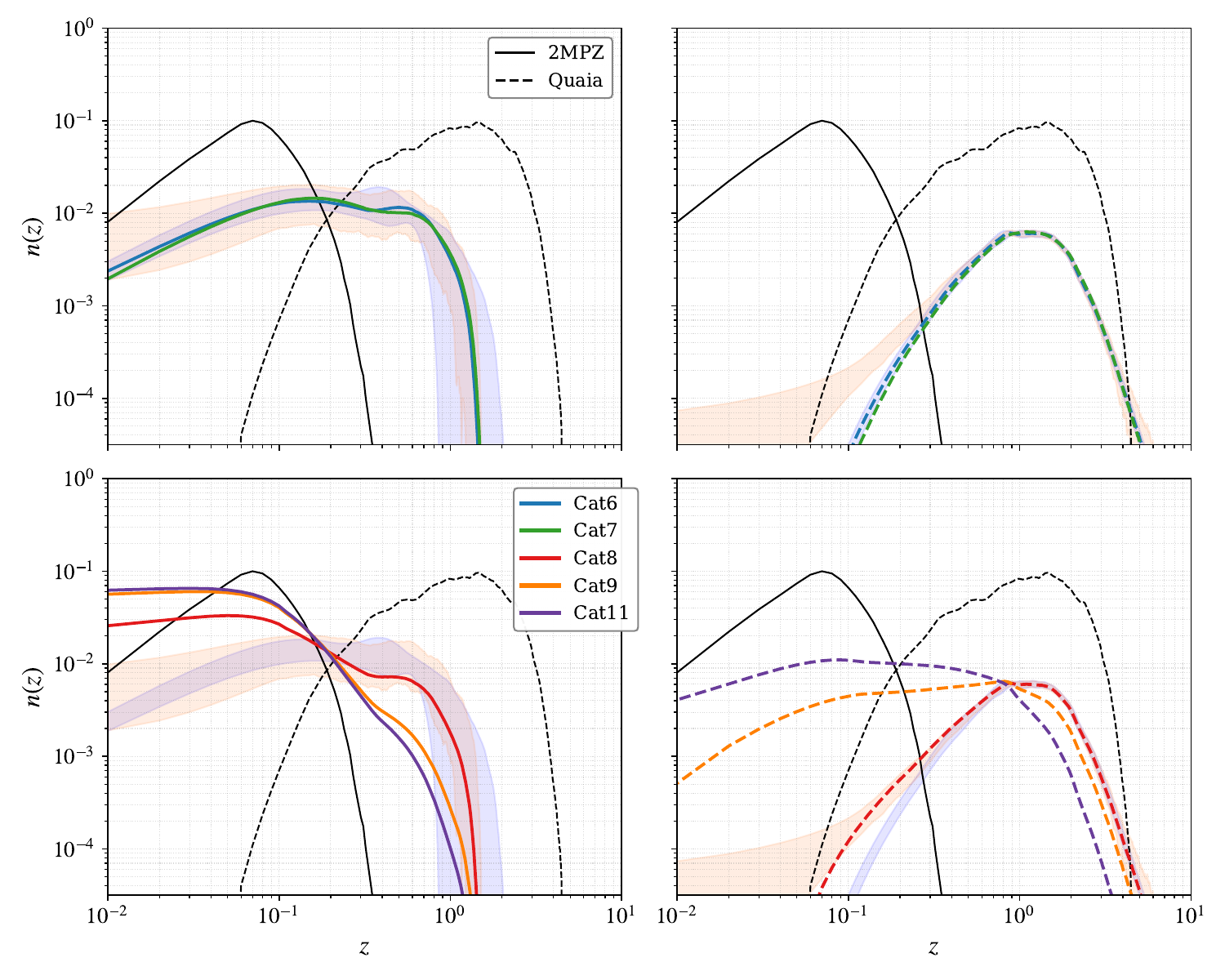}
    \caption{The GW radial kernel \(\phi_\mathrm{GW}(\chi)\) with the current-generation (left column, solid) and the next-generation (right column, dashed) detectors. The top panels compare catalogues~6 and~7, whereas the bottom panels compare catalogues~8, 9 and~11. The colour coding is the same as in \autoref{fig:rhodets}. The shaded light-blue areas show the range covered by applying the redshift errors of \autoref{fig:zerrs} to catalogue~6; the shaded light-orange areas show the sampling errors from the limited number of events in a given observation campaign. In addition, we show the 2MPZ and Quaia radial kernels in solid and dashed black, respectively. The $y$-axis units are arbitrary but consistent within tracers.}
    \label{fig:dets}
\end{figure}

We show the GW radial kernels for our five catalogues in \autoref{fig:dets}, alongside the redshift distribution of galaxies for the 2MPZ and Quaia galaxy catalogues (see \autoref{Sec:gal} for details on the latter two). In the left panels we show the expected GW radial kernels for current-generation detectors, whereas in the right panels we show the expected kernels for the next-generation. The top row shows the GW catalogues~6 and~7, which differ only in the choice of mass-transfer function, whereas the bottom row shows catalogues~8, 9 and~11, which differ in the choice of time delay. The multiple bumps clearly visible for current-generation catalogues~6 and~7, and slightly less pronounced for the other GW catalogues, come from the BHBH (lower redshift) and BHNS (higher redshift) events. These are not visible in the next-generation catalogues because these are expected to detect mostly NSNS and BHNS peaking at around the same redshift. From \autoref{fig:dets} we see that changing the time-delay exponent \(\alpha\) has a dramatic impact on the GW radial kernel: smaller values of \(\alpha\) squeeze the GW events to lower redshift, therefore changing the relative merger rates between current-generation and next-generation experiments.

In order to be able to determine the properties of the catalogues, and therefore the formation channel, the differences between radial kernels must be significant with respect to the uncertainty in the redshift determination. We find that the errors $\delta z$ of catalogue~6 completely encompass the kernel of catalogue~7 with both current-generation and next-generation detectors. Therefore, the nova mass-transfer function does not change the GW kernel in a significant way. If we want to be able to discriminate between the finer details of the mass-transfer function, accurate redshift information from, e.g.~an electromagnetic counterpart, or at least cross-identification of the host from a spectroscopic survey, is essential. This is all the more so if we include the statistical uncertainty from the limited Monte-Carlo sampling, given a real set of GW events, of the true underlying GW radial kernel. We show the size of this effect in \autoref{fig:dets}, displayed as an orange band; this is obtained by randomly selecting 15~instances of the expected 600~events for a year of current-generation observations (and $10^5$~events for a year of next-generation detectors) as applied to catalogue~6. We observe that, especially for current-generation detectors, the sampling uncertainty is roughly of the same order of magnitude as the redshift errors, and becomes dominant at low redshift (as expected because of the small number of events).

The situation is different however for catalogues~8, 9 and~11 which are generated with different time delays parametrised by $\alpha$. In particular, catalogues~9 and~11 can always be distinguished from catalogue~6. Catalogue~8, for which $\alpha = 0.95$, falls within the boundaries of the redshift uncertainty but only so for $z\gtrsim1$, whereas it remains distinct for closer objects. This shows that, because of the different redshift reaches of different generations of detectors, comparing the merger rates obtained from current-generation and next-generation detector networks can be utilised to understand finer details of the time delay, namely a 5\% difference in \(\alpha\), without the need for electromagnetic counterparts.

\subsubsection{GW beam factor}\label{Sec:beam}

The finite angular resolution of the GW detectors results in a sky localisation error of the GW signals on the 2-sphere. Ideally, because the sky localisation varies for each GW event in a given catalogue, as it depends on the network antenna pattern function and the event's SNR \(\rho\), we should run a Bayesian inference over the sky localisation parameters simultaneously with the dynamical parameters like masses and spins. For large mock populations this is computationally prohibitive and defeats the purpose of mock studies with synthetic GW catalogues. In~\citet{Calore:2020bpd} the sky localisation was numerically reconstructed using the fast reconstruction software \texttt{Bayestar} for each GW event and the resulting distribution fitted as a function of redshift \(z\);\footnote{\url{https://github.com/lpsinger/ligo.skymap}} the resulting \(\la\sigma(z)\ra_{\Delta z}\) (averaged in a given redshift bin or over the support of a galaxy catalogue) can be used to smear the GW field.

In this work we design a simpler and computationally cheaper method, which, as we will see, can be considered a good approximation to the \texttt{Bayestar} results. First of all, we assume that the sky localisation for each event can be expressed by a circular Gaussian beam of width \(\sigma\) (defined on the 2-sphere).\footnote{The localisation of GW events is in general not circular. Nonetheless, it can be proved that, especially for statistical analysis based on a population of GWs, the circular beam approximation is a very good description, see, e.g.~\citet{Calore:2020bpd}.} Second, we bin the GW catalogue into SNR bins of width~2.0 and assume events in an SNR bin to have a constant SNR in that bin, which we take to be the median. Third, we propose that the Gaussian width \(\sigma\) scales in inverse proportion to the SNR \(\rho\), namely \(\rho \propto 1/\sigma\). This is a reasonable approximation given that greater SNRs mean better evaluation of time-delays between the detectors and consequently better localisation, and it just follows from approximating the noise in the GW detector to be Gaussian. Proceeding in this way and normalising this inverse relationship from the GW170817 event, which was a loud event (\(\rho = 32.4\)) with \(\sigma\approx1.4\,\mathrm{deg}\), we compute the widths \(\sigma\) from the SNRs \(\rho\). Our approach produces $10^{-0.5}\leq\sigma\left[\mathrm{deg}\right]\leq 10^{0.5}$ for next-generation and $ 10^{-0.25}\leq\sigma\left[\mathrm{deg}\right]\leq 10^{0.75}$ for current-generation detectors, which we verified to closely approximate the $\sigma$ values to those obtained in~\citet{Calore:2020bpd}. Lastly, for each SNR bin, namely for each \(\sigma\), we build a GW field on the sphere with the events that fall into that SNR bin, and assign the angular beam factor
\begin{equation}\label{eq:beam1}
    \mathcal{B}^\mathrm{GW}_{\rho,\ell} = \sum_{j=0}^\ell \binom{\ell}{j} \binom{\ell+j}{j} \left(-\frac{\sigma(\rho)^2}{2}\right)^j \gamma\left(j+1, \frac{2}{\sigma(\rho)^2}\right) \,,
\end{equation}
where $\gamma(x,y)$ is the lower incomplete Gamma function, to these events. For small $\sigma$ and large $\ell$ the expression reduces to the usual form
\begin{equation}\label{eq:beam_s}
    \mathcal{B}^\mathrm{GW}_{\rho,\ell} \approx e^{-\frac{\ell(\ell+1)\sigma^2}{2}} \,.
\end{equation}

The binning in the SNR \(\rho\) not only affect the GW angular beam factor, but also the GW radial kernels $\phi(\chi)$ of \autoref{eq:ani}, because the redshift distribution of the GW events in each SNR bin only represents each particular bin. Therefore, in order to obtain the total XC we define a per-bin signal as
\begin{align}\label{eq:cl_limber-rho}
	\cS^\mathrm{GW\,g}_{\rho,\ell}& \deq \int \frac{\de\chi}{\chi^2}\,\phi_\mathrm{GW}(\chi,\rho) \,\phi_g(\chi) \, b_\mathrm{GW} \, b_\mathrm{g} \, P\left[z(\chi),k\right] \,,
\end{align}
and co-add them in proportion to the fraction \(w_\rho\) of events in each \(\rho\) bin:
\begin{align}\label{eq:cl_limber-add}
    \cS^\mathrm{GW\,g}_\ell & = \sum_\rho w_\rho \cS^\mathrm{GW\,g}_{\rho,\ell} \,.
\end{align}
The overall error on the XC is obtained from the covariance as
\begin{equation}\label{eq:nl}
    \left(\delta C^\mathrm{GW\,g}_\ell\right)^2 = \sum_{\rho,\,\rho'} w_\rho w_{\rho'} \mathsf{Cov}\left( \delta C^\mathrm{GW\,g}_{\rho,\ell} ,\, \delta C^\mathrm{GW\,g}_{\rho',\ell} \right) \,,
\end{equation}
where \(\delta C^\mathrm{GW\,g}_{\rho,\ell}\) takes into account the per-bin beams \(\mathcal{B}^\mathrm{GW}_{\rho,\ell}\), cf.~\autoref{eq:cross}. Lastly, as we have noted previously, while formally we should take into account the galaxy beam \(B^\mathrm{g}_\ell\), because the angular resolution of all galaxy catalogues considered in this work is vastly better than the GW ones, we can approximate \(\mathcal{B}^\mathrm{g}_\ell \approx 1\).

\subsection{Galaxies}\label{Sec:gal}

The anisotropy in galaxy number counts depends on the radial kernel of a putative galaxy catalogue \(\phi_{\rm g}(\chi)\), i.e.\ the weighted distribution of galaxy distances. The galaxy kernel is given by
\begin{align}\label{eq:g_ker}
	\phi_{\rm g}(\chi)\deq \left[\int \de\tilde\chi \, \tilde\chi^2\,\bar{n}_{\rm g,c}(\tilde\chi)\right]^{-1}\,\chi^2\,\bar{n}_{\rm g,c}(\chi) \;,
\end{align}
where \(\bar{n}_{\rm g,c}(\chi)=\chi^{-2}\,\de N_{\rm g}/\de\chi\) is the comoving (volumetric) number density of galaxies in the sample, \(\de N_{\rm g}\) being the (differential) angular number density of galaxies in a bin of radial width \(\de\chi\). In order to cover a broad range of redshifts and to capture the redshift evolution of the GW sources, in this work we model the galaxy catalogues on the specifics of two catalogues, the 2MASS Photometric Redshift catalogue 2MPZ~\citep{Bilicki:2013sza} and Quaia~\citep{Storey-Fisher:2023gca}. After removing all objects closer than $z = 0.005$ as we did for the GW events, the 2MPZ catalogue has a median redshift of $z_m = 0.081$ and 932,040 galaxies in total. For this catalogue we assume that a mask such as that described in~\citet{Koukoufilippas:2019ilu} will be employed in order to remove the galactic plane and other regions contaminated by stars or dust, which leaves a fraction $\fsky = 0.68$ of the sky unmasked. The Gaia-unWISE quasar catalogue (Quaia) is built from combining the Gaia quasar sample~\citep{Gaia} and the infrared data from unWISE~\citep{Meisner:2019lbf}. This catalogue contains 1,295,502 total objects and has a median redshift of $z_m = 1.5$. For this catalogue we assume a mask such as that defined in~\citet{Alonso:2023guh} will be used, thus leaving us with a $\fsky = 0.57$.\footnote{See also~\citet{Alonso:2024knf} for further details about how these catalogues are built.} We show the radial kernels of the galaxy catalogues in \autoref{fig:dets}.

\section{Results}\label{Sec:res}

\begin{figure}[htbp]
    \centering
    \includegraphics[width=\linewidth]{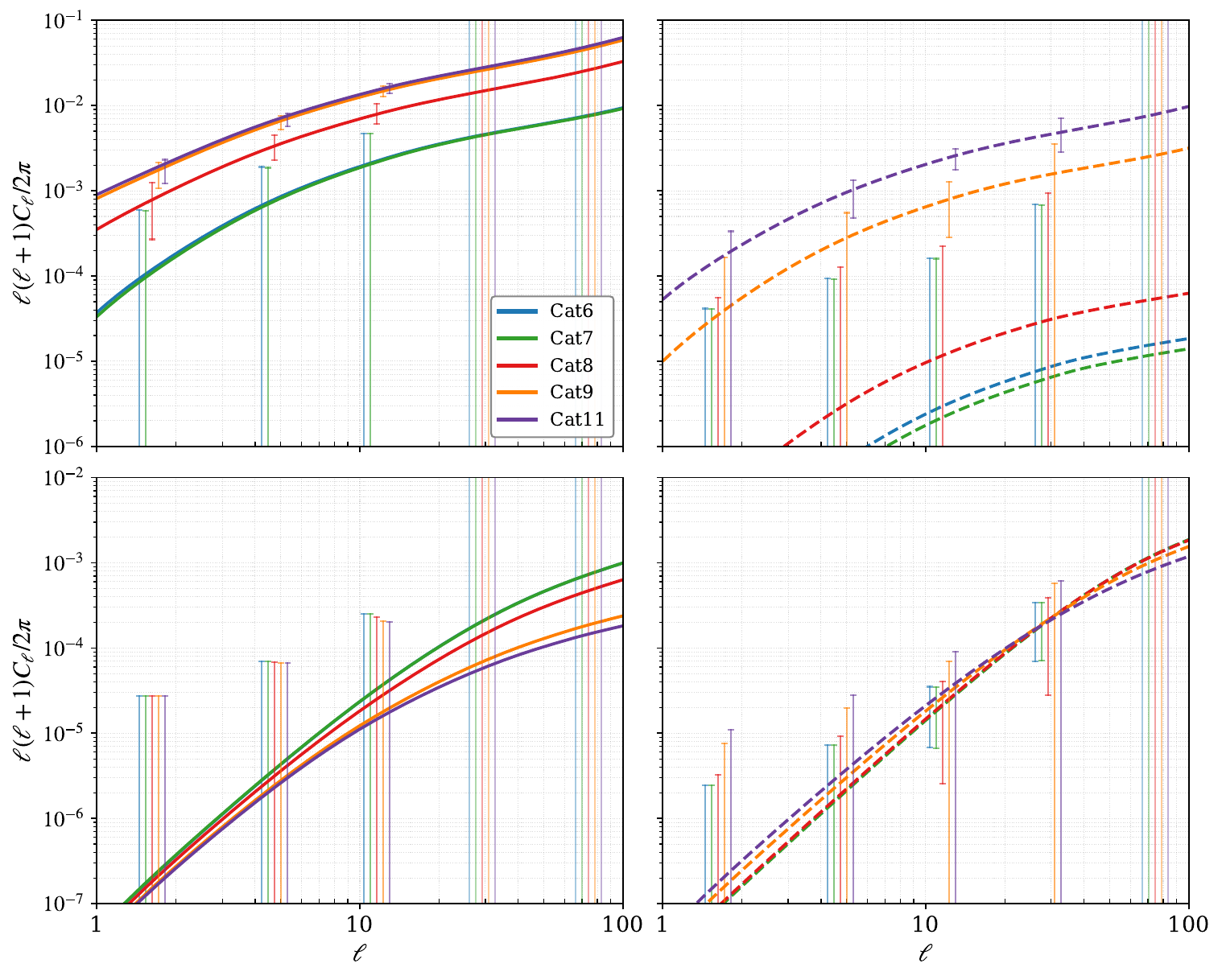}
    \caption{Gravitational wave-galaxy XC signal \(\cS_\ell\) for the five GW catalogues against the galaxy catalogues 2MPZ (top row) and Quaia (bottom row). On the left we show results for the current-generation detectors (solid lines), whereas the right are next-generation detectors (dashed lines). The error bars have been shifted along the x-axis for clarity.}
    \label{fig:GWGal}
\end{figure}

In \autoref{fig:GWGal} we show the gravitational wave-galaxy XC signal \(\cS^\mathrm{GW\,g}_\ell \deq \cS_\ell\) (where we drop the superscript to avoid clutter) for the five GW catalogues~6, 7, 8, 9 and~11 (recall that we have assumed the detection threshold SNR values of $\rho\geq8$ for HLV and $\rho\geq12$ for ET2CE). In order to obtain the XC we have binned the \(\ell\) space in five equal-size logarithmic bins from \(\ell=1\) to \(\ell=100\) -- given the limited angular resolution of the GW events we expect to have no power beyond this scale, as it is confirmed by the large error bars in the last bandpower. The top row shows the XC of the GW catalogues with 2MPZ galaxies, whereas the bottom row is for Quaia. The left column (solid lines) corresponds to current-generation GW detectors, the right column (dashed lines) corresponds to next-generation GW detectors.

First we observe that, for current-generation GW detectors, the XCs with the shallow 2MPZ catalogue are significantly different depending on the merger history, and, in particular at medium scales \(\ell\sim{\cal O}(10)\), it is possible to distinguish between time-delay functions \(1/\tau\) and \(1/\tau^{0.95}\) and between the latter and the much longer delays of catalogues~9 and~11 (but it is not possible to distinguish~9 from~11). This is evident from how the time-delay functions change the radial kernels, with longer delays boosting low-redshift end and therefore significantly improving the overlap with the 2MPZ catalogue. The cumulative signal-to-noise ratios for these XCs range from \(1\) (i.e.\ non-detection) for catalogues~6 and~7, up to about \(10\) for catalogues~9 and~11. Conversely, the overlap between all GW catalogues and Quaia is mostly unaffected by the merger history, and all the XCs are virtually indistinguishable from each other; in terms of detection significance, the current-generation Quaia XCs are all compatible with zero.

For next-generation GW detectors the overlap between the detected GW events and the 2MPZ galaxies is much reduced, the more so for shorter time delays corresponding to catalogues~6, 7 and~8, whose XC signals are indistinguishable from each other as well as overall undetectable. Longer time-delay functions boost the signal of the XCs; at medium angular scales we observe that the time delays with \(\alpha=0.75\) (catalogue~9) and \(\alpha=0.50\) (catalogue~11) are clearly separate. Their cumulative signal-to-noise ratios are of order \(2\) and \(5\), respectively, underscoring how the XC can be confidently detected only for a GW population generated with the longest time delays.

Next-generation GW catalogues all exhibit a very similar radial kernel overlap with the deep catalogue Gaia; similarly to current-generation GW detectors, the XC signals for all five galaxy catalogues are very similar and their differences well within the error bars. Nonetheless, the short time-delay function catalogues~6 and~7, even though indistinguishable from each other, are marginally detectable (cumulative signal-to-noise ratio of about \(2\)) whereas the remaining XCs are undetectable.

Lastly, as we already observed, changing the parameters of the formation channel also changes the ratio between current-generation to next-generation total detections. This is visible in the relative ratios between XC signals obtained with current and future detectors, with 2MPZ correlations changing by up to over three orders of magnitude for catalogues~6, 7 and~8. If a detection of the XC can be made then, by comparing current-generation and next-generation detections, we could in principle be able to say something about specific formation channels.

\section{Conclusion and outlook}\label{Sec:out}

In this work we analysed the impact of the history of binary formation and evolution until merger on the harmonic, angular cross-correlation between GW events and the large-scale structure as traced by galaxies. To characterise the GW populations we have focussed on two parameters: the progenitor-to-remnant mass-transfer function, rapid or delayed nova, and the probability distribution function for the time delay \(\tau\) between progenitor and remnant, which we allowed to vary between \(0.5\) and \(1\), with the latter being the most common choice in the literature. The GW population distribution in the sky is further determined by the detection threshold SNRs (we set 8 for current-generation and 12 for next-generation facilities) and the precision of the localisation, which we tie to the detection SNR with inverse proportionality. We therefore produce five GW catalogues, normalised to the expected HLV O5 detection rate, which we cross-correlate with two galaxy catalogues modelled on the 2MPZ (low-redshift) and Quaia (high-redshift) samples.

We find that the merger history has a pronounced impact on the XC signals at all scales for the 2MPZ-like catalogue, with differences of up to three orders of magnitude between time delay functions \(1/\tau\) and \(1/\tau^{0.5}\) for next-generation facilities, owing to their ability to reach deeper in redshift, which in turn makes these detectors more sensitive to changes in the GW kernels. Even with current-generation facilities the differences between the XC obtained with the fastest and slowest progenitor-to-remnant time-delay functions is over an order of magnitude. In all cases, the differences for a very deep catalogue such as Quaia are much less pronounced because our GW catalogues differ most at low redshift.

The \(\alpha=0.75\) and \(\alpha=0.5\) GW populations for current-generation facilities (catalogues~9 and~11) when cross-correlated with the 2MPZ galaxies would be very easily detected but can not be told apart. The XC of the next-generation \(\alpha=0.5\) GW population (catalogue~11) with 2MPZ can also be detected, whereas all other merger histories would produce GW events whose XCs are not significant. Lastly, we find that all the XCs with Quaia are at best marginally detectable and indistinguishable from each other. This means that, in the simplified case that we have analysed, using the XC alone and assuming detection, it will not be possible to distinguish different formation channels with the exception of the most extreme cases with a very long time-delay function.

Our exploratory results point to the importance of the formation channel in the determination of the angular, harmonic XC with galaxies.  Whilst we restricted our investigation to two parameters that regulate the GW radial kernel, changes in the trapping factor, the star-formation rate, the initial mass function as well as the host metallicity could also produce very different predictions for the XC. Because the use of the XC as a tool to determine astrophysical or cosmological parameters, e.g.\ the GW-galaxy bias or the present value of the Hubble rate, hinges on the significance of the detection of the XC itself, the whole merger history can have a profound impact on any quantity derived from the XC.

\section*{Acknowledgements}

The authors would like to thank Francesca Calore and Stefano Camera for constructive feedback on the manuscript. KC acknowledges research support by the PPLZ grant of the Czech Academy of Sciences (Project No.\ 10005320/0501). FU acknowledges support from the European Structural and Investment Funds and the Czech Ministry of Education, Youth and Sports (project No.\ FORTE--CZ.02.01.01/00/22\_008/0004632). The authors acknowledge use of the local CEICO clusters \emph{Phoebe} and \emph{Koios}, as well as the \emph{Metacentrum} computing facilities.\footnote{\url{https://www.metacentrum.cz}} Cluster support for the runs was provided by the CEICO HPC wizard Josef Dvo\v{r}\'{a}\v{c}ek.

\bibliographystyle{aasjournal}
\bibliography{ref}

\end{document}